\documentclass[11pt]{article}
\usepackage[margin=1in]{geometry}
\usepackage{graphicx}
\usepackage{xspace}
\usepackage{amsmath}
\usepackage{wrapfig}

\newcommand{\nonu}{0$\nu\beta\beta$\xspace}
\newcommand{\mbb}{$\langle m_{\beta\beta}\rangle$\xspace}

\def\beq{\begin{equation}}
\def\eeq#1{\label{#1}\end{equation}}
\def\eeqn{\end{equation}}

\def\beqa{\begin{eqnarray}}
\def\eeqa#1{\label{#1}\end{eqnarray}}
\def\eeqan{\end{eqnarray}}

\let\bar=\overbar

\def\Dslash{\not{\hbox{\kern-4pt $D$}}}
\def\dslash{\not{\hbox{\kern-2pt $\del$}}}

\def\msb{{\bar{\ssstyle M \kern -1pt S}}}

\def\Title#1{\begin{center} {\Large {\bf #1} } \end{center}}
\def\Author#1{\begin{center} {\normalsize {\sc #1} } \end{center}}
\def\Institution#1{\begin{center} {\normalsize {\it #1} } \end{center}}
\def\Abstract#1{\noindent {\normalsize {\bf Abstract:} {\normalfont #1}}}
\def\Conference{\vspace{4mm}\begin{raggedright} {\normalsize {\it Talk presented at the 2019 Meeting of the Division of Particles and Fields of the American Physical Society (DPF2019), July 29--August 2, 2019, Northeastern University, Boston, C1907293.} } \end{raggedright}\vspace{4mm}}

\begin{document}

%%%%%%%%%%%%%%%%%%%%%%%%%%%%%%%%%%%%%%%%%%%%%%%%%%%%%%%%%%%%%%%%%%%%%%%%%%%
%
% TITLE, AUTHOR, INSTITUTION, ABSTRACT ==> UPDATE
% 
%%%%%%%%%%%%%%%%%%%%%%%%%%%%%%%%%%%%%%%%%%%%%%%%%%%%%%%%%%%%%%%%%%%%%%%%%%%

\Title{Nano-tracking detector for neutrinoless double beta decay characterization}

\Author{Ethan Brown, Kelly Odgers, Adam Tidball}

\Institution{Department of Physics, Applied Physics and Astronomy\\ Rensselaer Polytechnic Institute, Troy, NY, USA}

\Abstract{Of the many extensions to the standard model that could possibly generate neutrino mass, most necessitate the neutrino being a Majorana fermion. If this is the case, the rare process of neutrinoless double beta decay is predicted with half lives greater than about $10^{25}$ years. Many current and future experiments look for this decay by identifying a summed double beta energy at the Q value of the decay, but adding energy and angular measurements of the individual betas allows the underlying decay mechanism to be probed. %\\
A novel nano-tracking detector based on a clever combination of thin film CdTe devices will be presented here. This tracker will have order 100 nm spatial resolution in one dimension while measuring the energy deposition across the track length of electron recoils in the detector. This allows energy and angular correlation measurements of a potential neutrinoless double beta decay signal, as well as a unique background suppression capability. Deep learning algorithms will be used to reconstruct the electron paths, the double beta signals and perform the correlation analyses, and will simultaneously allow clear distinction between double betas and single beta or gamma-induced electronic recoils. %\\
The detector concept will be presented, along with preliminary studies, to demonstrate its operation and the physics reach for neutrinoless double beta decay. By exploiting recoil discrimination, an array of these detectors can potentially probe beyond the inverted hierarchy to either follow the next generation of neutrinoless double beta decay experiments or to serve as a post-discovery characterization experiment.}

\Conference

%%%%%%%%%%%%%%%%%%%%%%%%%%%%%%%%%%%%%%%%%%%%%%%%%%%%%%%%%%%%%%%%%%%%%%%%%%%
%
% MAIN TEXT ==> UPDATE
% 
%%%%%%%%%%%%%%%%%%%%%%%%%%%%%%%%%%%%%%%%%%%%%%%%%%%%%%%%%%%%%%%%%%%%%%%%%%%

\section{Introduction}
Nano-scale particle tracking would allow measurements of the kinetics of nuclear decay products and low energy particles and beams. Scaling down the tracking concepts used in high energy physics to 100 nm or less has widespread applications, including x-ray imaging, electron beam imaging (SEM, TEM, etc), and radiodecay characterization. The new detector concept presented here would provide $\sim$100 nm resolution in a 2D tracker, and application to characterization of neutrinoless double beta (\nonu) decay.

Extensions to the standard model (SM) of particle physics that account for non-zero neutrino mass often require Majorana neutrinos. Such models predict the rare second order process called (\nonu) decay, where two electrons are emitted from a decaying nucleus without accompanying antineutrinos. This violates lepton number conservation, a yet to be observed process that is needed to explain the matter-antimatter asymmetry in the universe. Its observation would provide an indirect measurement of the neutrino mass. In the simplest case of a light Majorana neutrino exchange, the half life is related to an average effective neutrino mass \mbb by,
\begin{equation}
    \big[T_{1/2}^{0\nu}\big]^{-1} = G|M|^2  \langle m_{\beta\beta}\rangle^2,
\label{eqn:thalf}
\end{equation}
where $G$ is the phase space factor and $M$ is the nuclear matrix element.

The experimental search for \nonu decay looks for a summed energy of the two betas at the Q value of the decay, as only a negligible amount of energy is imparted to the nucleus. These searches use several hundred kg of $^{136}$Xe, $^{76}$Ge, or $^{130}$Te, where lack of a conclusive \nonu decay signal has allowed limits to be set, ranging from $10^{25} - 10^{26}$ years \cite{bib:GerdaLimit, bib:KamlandZenLimit, bib:EXO200FinalResults, bib:CuoreLimit}. Future experiments will be sensitive to half lives as large as $10^{28}$ years, and could well discover \nonu decay \cite{bib:nEXOSensitivity}.

While observing \nonu decay would answer important questions about the nature of the neutrino and extensions to the SM, there is much that could be learned from the angular and energy distributions of the betas in the decay. For example, left-right symmetric models of neutrinos add a natural symmetry between the left and right handed states by introducing a seesaw with a heavy right handed partner for each SM neutrino, as well as a helicity-inverted copy. The interaction eigenstates are then linear combinations of the light and heavy states, and the PMNS matrix is then generalized as,
\begin{equation}
    %\begin{multline}
    \begin{split}
    \nu_{eL}' = \sum_k^{light} U_{ek} \nu_{kL} +
    \sum_k^{heavy} S_{ek} N_{kR}^c\\
    %\end{center}
    %\begin{center}
    \nu_{eR}' = \sum_k^{heavy} T^*_{ek}\nu_{kL}^c +
    \sum_k^{light} V^*_{ek} N_{kR},
    \end{split}
    %\end{multline}
    \label{eqn:NuMix}
\end{equation}
where $\nu_{kL}$ are the standard left handed light neutrino mass eigenstates, $N_{kR}$ are the heavy right handed states, $\nu_{kR}$ are the light right handed states, and $N_kR$ are the heavy right handed states. The mixing matrix then factors into $U_{ek}$ (the PMNS matrix), $S_{ek}$, $T_{ek}$, and $V_{ek}$.

In such models, additional terms are present in the matrix element in equation \ref{eqn:thalf}. These can be probed by measurements of \nonu decay rates in multiple isotopes and by measuring energy and angular correlations of the emitted betas. Thus, a detector based on multiple isotopes with the ability to resolve the individual betas is ideal for a post-discovery characterization of \nonu decay. Such an approach is considered for SuperNEMO using thin foil \nonu decay sources \cite{bib:SuperNEMO}, and here we describe a new concept where the detector is constructed of the source material.

\section{Detector concept}
A new type of 2D tracking calorimeter is described here. It will use PIN diodes comprised of intrinsic crystaline CdTe between p and n doped layers \cite{bib:PINDiode}. Incident radiation or a recoil from radiation excites electron-hole pairs to the conduction band, which are then drifted to the p and n contacts by the reverse biased field \cite{bib:FilmGammaDetector}. 

\begin{wrapfigure}{r}{.55\linewidth}
    %\vspace{-6pt}
    \centering
    \includegraphics[width=\linewidth]{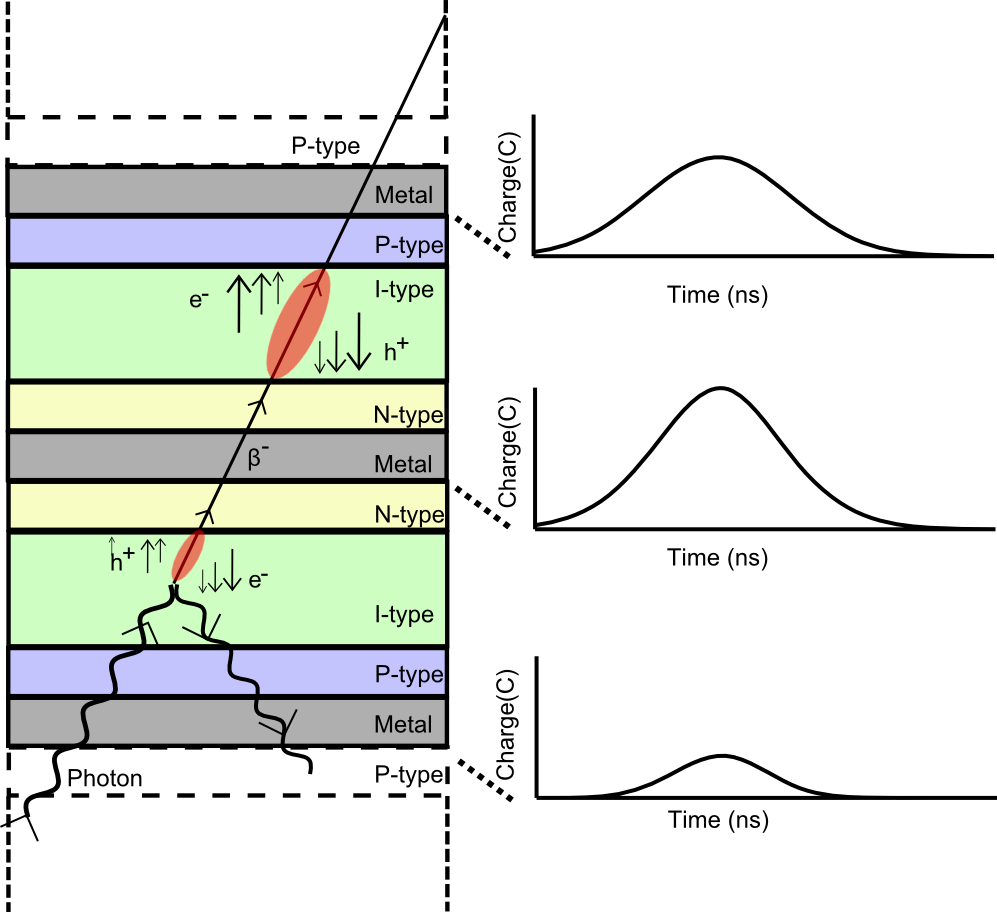}
    \caption{Schematic of detector concept. Each layer consists of a $\sim$100 nm thick PIN diode. Alternating the orientation allows segmented biasing and readout of avalanche signals. Shown here is a single Compton induced electron recoil.}
    \label{fig:detector}
    \vspace{-12pt}
\end{wrapfigure}

A schematic of the detector is shown in figure \ref{fig:detector}. PIN diodes can be constructed with an intrinsic layer thickness on the order of tens to hundreds of nm, with order 10 nm P and N layers on either end as well as an order 10 nm conductive layer for signal readout and diode biasing. These can then be stacked with an alternating orientation (PIN - NIP - PIN...) to yield a multi-layered tracker. As shown for the case of an electron recoil from Compton scattering, when an incident gamma scatters in the detector, the recoiling electron will liberate electron-hole pairs which will avalanche as they propagate to the conductive layers, where the current can be read. The current amplitude on each channel scales with the energy deposition in the adjacent I layers allowing a segmented calorimetric measurement. Thus, as the track length within a single channel depends on the angle of the recoiling electron, that angle can be determined based on the energy loss in each layer, allowing nano-scale resolution while simultaneously measuring the energy.

Two dimensional tracking information is possible by localizing the depth of each channel within the detector and reconstructing the polar angle based on the energy loss per unit distance. Normally incident tracks will traverse channels with minimum length while glancing angles will travel much farther through each layer. 

While the concept of a layered semiconductor detector is not new, previous work used layers hundreds of $\mu$m thick, three orders of magnitude larger than what is considered here \cite{bib:StackedPatent}. Similar ideas were even applied for radiation calorimetry \cite{bib:RadPatent}, \cite{bib:SCRadPatent}, but this is the first time this concept will be combined with calorimetry for the purpose of tracking. This work also builds on the existing concept of nano-tracking, as commercial graphene FETs have been proposed for pixelated trackers for nuclear recoil detection \cite{bib:Ptolemy}. However, while graphene pixels suffer from large dead regions between the active graphene sites due to the large electrode support structure surrounding a single monolayer of graphene \cite{bib:GrapheneTransistor}, the nano-tracking detector presented here is built predominantly of active material in the intrinsic layers of the diodes, which is crucial for achieving good energy resolution.

\begin{wrapfigure}{r}{.3\linewidth}
    \vspace{-13pt}
    \includegraphics[width=\linewidth]{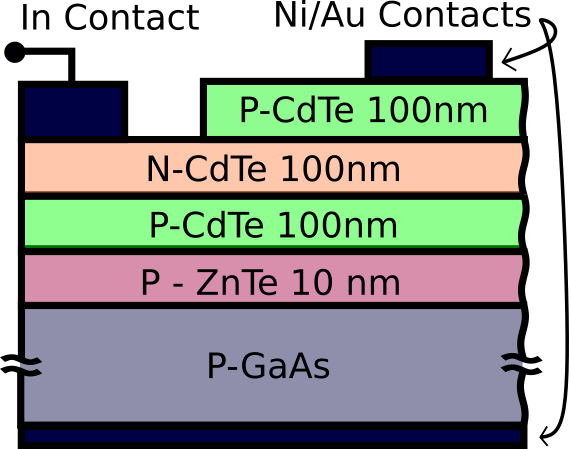}
    \caption{Example of layered diode construction mechanism (for PN diodes here) based on MOCVD deposition.}
    \label{fig:Layers}
    \vspace{-12pt}
\end{wrapfigure}
The choice of CdTe for the diodes is driven by a combination of \nonu decay physics and feasibility of device construction. Natural Cd and Te each have two double beta emitting isotopes, $^{114}$Cd, $^{116}$Cd, $^{128}$Te, and $^{132}$Te. Isotopic separation may be necessary to avoid swamping the \nonu decay signal with the two neutrino decays of higher Q value isotopes. But using 4 isotopes allows multi-isotope characterization of decays for model probing.

Additionally, CdTe devices are widely developed for light and radiation detection \cite{bib:SolarCell, bib:Bhat1, bib:Bhat2}. Epitaxial growth of large crystals is routine using metal-organic chemical vapor deposition (MOCVD) \cite{bib:Bhat-3}. An example of prototype development of a nanotracker with stacked diodes is shown in figure \ref{fig:Layers}. This shows a mechanism for constructing PN junctions for each layer. A similar mechanism can be used for PIN diodes. Here, there are three electrodes for readout of two diodes, the P-ZnTe layer, and the In and Ni/Au contacts. This prototype development mechanism is based on previous production of CdTe diode structures. Thus, the detector concept presented here is built around the feasibility of developing prototypes.

\section{Tracking and calorimetry}
Track and energy information can be extracted from the energy deposition in each channel. Since each channel is short relative to the range of $\sim$MeV electrons, betas from \nonu decay will cross hundreds of channels. By summing the energy in all channels, the total recoil energy can be measured. And since glancing angle betas will traverse much fewer channels than those with normal incidence, the number of channels and energy density in each channel can provide the polar angle. 
% \begin{wrapfigure}{r}{.65\linewidth}
%     %\vspace{-12pt}
%     \centering
%     \includegraphics[width=\linewidth]{figures/Events.png}
%     \caption{Simulated waveforms for normal incidence events with (top) two anti-parallel 1.2 MeV electrons, (middle) one 1.2 MeV electron, and (bottom) one 2.4 MeV electron.}
%     \label{fig:Events}
%     \vspace{-12pt}
% \end{wrapfigure}
Additionally, the Bragg peak allows identification of the end of individual tracks, allowing determination of the direction and multiplicity of an event. Although this detector has $\sim$10\% dead volume, since radiation tracks extend over tens to hundreds of channels, fluctuations in energy deposition in each dead volume can be averaged out, and an energy resolution on the order of percent is likely possible.

\begin{figure}[!h]
    \centering
    %\begin{subfigure}%{.45\linewidth}
    \includegraphics[width=.48\linewidth]{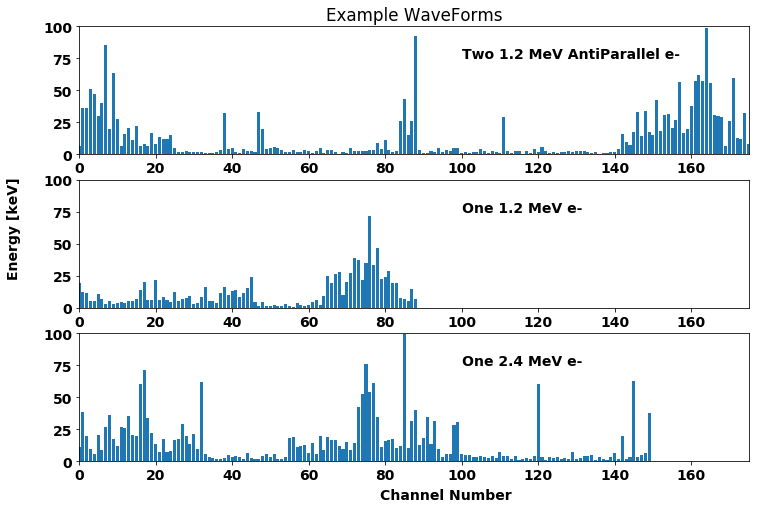}
    %\end{subfigure}
    %\begin{subfigure}%{.45\linewidth}
    \includegraphics[width=.48\linewidth]{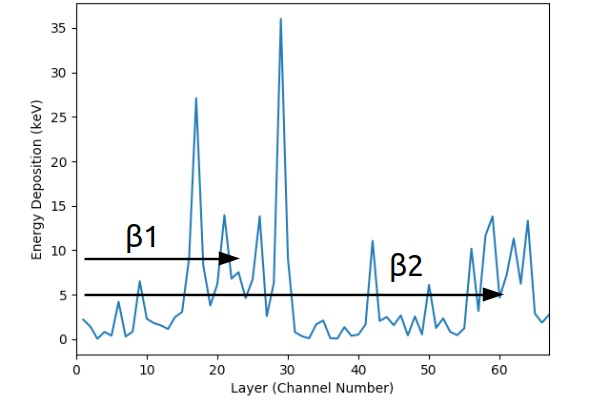}
    %\end{subfigure}
    \caption{(Left) simulated waveforms for normal incidence events with (top) two anti-parallel 1.2 MeV electrons, (middle) one 1.2 MeV electron, and (bottom) one 2.4 MeV electron. (Right) \nonu decay event with isotropic angles for both betas.}
    \label{fig:Events}
\end{figure}

% \begin{wrapfigure}{r}{.5\linewidth}
%     \vspace{-12pt}
%     \centering
%     \includegraphics[width=\linewidth]{figures/0vBBEvent.png}
%     \caption{\nonu decay event with isotropic angles for both betas}
%     \label{fig:0vBBEvent}
%     \vspace{-12pt}
% \end{wrapfigure}
\begin{wrapfigure}{r}{.5\linewidth}
    \vspace{-12pt}
    \centering
    \includegraphics[width=\linewidth]{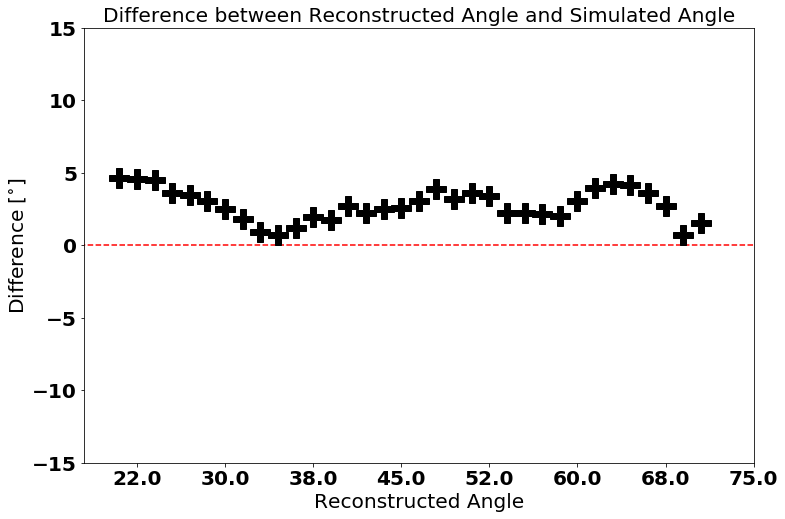}
    \caption{Example of angular reconstruction of single 1 MeV recoiling electrons using a neural net.}
    \label{fig:Reconstruction}
    \vspace{-12pt}
\end{wrapfigure}
Simulations were conducted in Geant4 to generate example waveforms. The simulation includes the active CdTe, as well as dead layers for electrodes and depletion layers in the diodes. The detector was constructed using 1000 layers with a thickness of 100 nm, and a dead layer thickness of 10 nm. Figure \ref{fig:Events} left shows three distinct cases of normally incident electrons, with two antiparallel 1.2 MeV electrons, one 1.2 MeV electron, and one 2.4 MeV electron. The Bragg peaks are clearly visible at either end in the top panel, near the middle of the center panel, and at the right edge of the lower panel. This demonstrates qualitatively how these events can be distinguished.

An example of a more complex \nonu decay event is shown in figure \ref{fig:Events} right, where the two betas have arbitrary direction relative to one another and to the detector normal. Here the Bragg peak of each beta is still clearly visible, and the vertex of the \nonu decay is evident (labeled channel 0 here). But to determine the best estimate of the angle and energy, a more sophisticated analysis is needed. 

Machine learning algorithms have been developed, such as deep neural nets, to optimize the energy and angular reconstruction. An example of the angular reconstruction is shown in figure \ref{fig:Reconstruction} for the case of single 1 MeV electron recoils. This yields few degree precision in reconstruction. Performance is expected to improve by using a deep neural net that is tuned to identifying specific features in the data. Additionally, fine tuning the detector geometry is also expected to improve performance.

\section{Application for \nonu decay}
Three features of the detector concept make it ideal for searches and characterization of \nonu decay. The tracking ability allows angular correlation measurements of a potential signal, while the calorimetry provides isolation of double beta signals at the Q value (where the sum of both beta energies equals the Q value). Additionally, the ability to identify recoil types and distinguish single and double electrons provides an unprecedented background discrimination potential. 

\begin{wrapfigure}{r}{.5\linewidth}
    \vspace{-12pt}
    \centering
    \includegraphics[width=\linewidth]{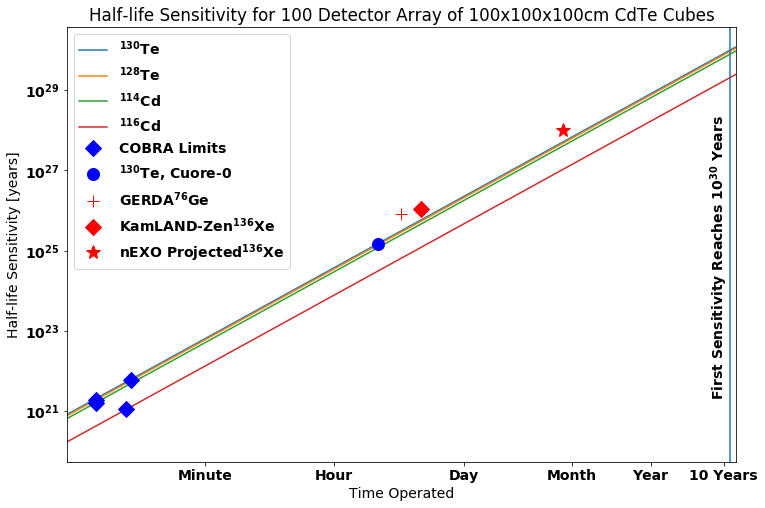}
    \caption{Example of sensitivity vs time for an array of 100 detectors with dimensions of order 100 cm.}
    \label{fig:Sensitivity}
    \vspace{-12pt}
\end{wrapfigure}
The challenge of observing \nonu decay is the huge amount of double beta emitting isotopes required. In order to realize this in a nano device, macroscopic detectors must be constructed. While constructing detectors with dimensions large than a few cm poses significant challenges, if dimensions can be pushed to 100cm, an array of 100 detectors would provide enough target mass to reach a half life sensitivity on the order of $10^{30}$ years with a few years of data. As shown in figure \ref{fig:Sensitivity}, this is well beyond the target of the next generation of \nonu decay experiments, and is poised to continue the search in the normal hierarchy should \nonu decay not be found, or to conduct high statistics measurements as a post discovery experiment.

\section{Conclusion}
We describe here a novel nano-tracking detector capable of resolving 2D track information of nuclear decay products, and its application for characterization measurements of \nonu decay. A clever assembly of $\sim$100 nm thick CdTe PIN diodes can be used to measure the polar angle and energy of individual betas from the decay. It provides identification of event multiplicity and a unique background discrimination capability, and the deployment of an array of large scale detectors would provide sensitivity to the \nonu decay half life at the level of $10^{30}$ years. 

The detector concept has been developed around a feasible construction mechanism, and prototype production is now ready to begin. In addition to \nonu decay, this detector can be used in vast applications of radiation detection where $\sim$100 nm spatial information is required.

\section*{Acknowledgements}
We gratefully acknowledge Ishwara Bhat for consultation on the detector design and conceptual development for prototyping.

\bibliographystyle{IEEEtran}
\bibliography{refs.bib}
 
\end{document}